\documentclass[twocolumn,twoside,slac_two]{revtex4}
\usepackage{graphicx}
\usepackage{fancyhdr}

\usepackage{amsmath}

\begin{document}

\title{QCD Factorization and SCET at NLO}

%

\author{Alexander R. Williamson}
\affiliation{Department of Physics, Carnegie Mellon University, Pittsburgh, PA 15213}

\begin{abstract}
In this article we review recent advances in the area of QCD factorization.
We begin with a brief outline of QCD factorization, as well as a description of some recent results regarding $B \to VV$ decays.  We examine what is necessary at NLO, look at recent advances towards that goal and consider what remains to be done.
\end{abstract}

\maketitle

\thispagestyle{fancy}


\section{Introduction}

Charmless hadronic two-body $B$ decays are a rich source of flavour physics information and are particularly well suited to the study of $CP$ violation.  The theoretical description of these decays is hampered by the need to separate the long distance physics of hadronization from the short distance electroweak physics of interest.  QCD factorization (QCDF) \cite{hep-ph/9905312,hep-ph/0006124,hep-ph/0104110} is a theoretical framework suitable for the study of these decays.  At leading order (LO) in $1/m_b$, it allows for a clean separation of scales; the long distance physics is parametrized in terms of form factors and decay constants which can be related to other processes, while the short distance physics, including strong phases, is perturbatively calculable.  The effective field theory formalism necessary to go beyond LO in $1/m_b$ is the Soft Collinear Effective Theory (SCET) developed 
by Bauer, Fleming, Pirjol and Stewart. \cite{hep-ph/0401188}  

QCDF was supplemented by an effective field theory description, SCET, giving further credibility to the factorization result first put forward in the context of QCDF.  A modern view of QCDF, that we adopt in this short review on the status of QCDF, would be that QCDF is the application of SCET to two-body B decays.  We therefore use the terms QCDF and SCET interchangeably throughout the review.  In the application of the theory, two approaches exist: the original approach of Beneke, Buchalla, Neubert and Sachrajda (BBNS), often called QCD factorization (QCDF) and the approach advocated by Bauer, Pirjol, Rothstein and Stewart (BPRS) often referred to as SCET.
Although this article will make reference to both approaches, emphasis will be placed on that of BBNS.  The discussion reflecting recent results from the BPRS perspective has been covered at this conference by Christian Bauer in his presentation on SCET.

In the next section we give a basic description of charmless two-body hadronic $B$ decays in the QCD factorization formalism and describe the differences which exist between the BBNS and BPRS approaches.  In Section \ref{BVV}, we discuss a recent application of QCD factorization to $B \to VV$ decays.  In Section \ref{NLO}, we discuss the theoretical situation at NLO, including both $1/m_b$ and $\alpha_s$ corrections, some recent work towards the NLO perturbative corrections and what work remains to be done.

\section{The Basics of Factorization}

At LO in $1/m_b$, the amplitude for $B \to M_1 M_2$ decay factorizes,
\begin{eqnarray}
\label{LOeqn}
\langle M_1 M_2|O|\bar B\rangle &=& F^{B\to M_1}T^\mathrm{I}*f_{M_2} \Phi_{M_2} + M_1 \leftrightarrow M_2 \nonumber \\
&\,& \hspace{-2em}+ T^\mathrm{II} *f_{B} \Phi_{B}*f_{M_1} \Phi_{M_1}*f_{M_2} \Phi_{M_2}.
\end{eqnarray}
$T^\mathrm{I}$ and $T^\mathrm{II}$ are hard scattering kernels that depend on the specific process
being considered, but are perturbatively calculable in an expansion in $\alpha_s(m_b)$ and $\alpha_s(\mu_i)$, where $\mu_i$ is the intermediate scale $\mu_i=\sqrt{\Lambda_\mathrm{QCD}m_b}$.
$f_M$ and $f_B$ are light and heavy decay constants, while $\Phi_M$ and $\Phi_B$ are light cone distribution amplitudes (LCDA).  They are universal and do not depend on the decay process.  They are determined from QCD sum rules, lattice QCD, or experiment.  Although this formula is valid only at LO in $1/m_b$, it is valid to all orders in $\alpha_s$.  A systematic treatment of the higher order corrections can be given using SCET.

\subsection {The BBNS and BPRS Approaches}

As discussed in the introduction, there are two main approaches which we refer to as BBNS
\cite{hep-ph/9905312,hep-ph/0006124,hep-ph/0104110}
and BPRS \cite{hep-ph/0401188}.
Below, we outline several of the important differences, which have been discussed at great length in \cite{hep-ph/0411171, hep-ph/0502094}.

\begin{figure}
\caption{\label{CP}A typical charming penguin contribution}
\includegraphics{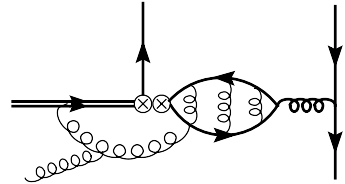}
\end{figure}

Perhaps the most important difference between the two approaches is the treatment of the so-called charming penguin contributions (see Fig. \ref{CP}).   BPRS argue that since $2m_c\sim m_b$, there are configurations with almost on-shell charm quarks leading to long distance effects.  As a result, they parametrize the penguins and fit them from data.  
BBNS maintains that charming penguin diagrams factorize in the formal $1/m_b$ limit and should not be given special treatment.

In the BBNS approach a subset of higher order $1/m_b$ corrections are typically included that are deemed to be numerically large.  This is in contrast to BPRS, where calculations are consistently performed to a fixed order in the $1/m_b$ expansion.

It has been the philosophy of BBNS that the inputs used to make predictions for $B \to M_1 M_2$ decays do not come from these decays themselves; rather, the form factors and decay constants enter from other observables, lattice calculations, or some other external source.  BPRS on the other hand impose no such restriction and use not only external sources but also the $B\to M_1 M_2$ decays themselves as inputs.
In particular, BPRS has up to this point chosen to fit the SCET form factor, $\Xi$ (see below), rather than perturbatively expanding it and relating it to other observables.

BBNS and BPRS also differ in the counting of the soft overlap and hard scattering terms in (\ref{LOeqn}).
In both cases these terms are counted as leading order in $1/m_b$, but
BBNS assumes that soft overlap is dominated by soft contributions without $\alpha_s(\mu_i)$ suppression, and is thus assumed to be larger.

The differences given above leave the BBNS approach more predictive than BPRS, but at the cost of being less conservative. 
We will see that when NLO corrections are taken into account, some of these differences will disappear.

\section{\label{BVV} Recent Results for $B \to VV$}

QCDF had been used to predict the branching ratios and $CP$ asymmetries of a large number of two-body charmless $B \to PP$ and $B \to PV$ decay modes
\cite{hep-ph/0210085, hep-ph/0308039}.
In a recent work, Beneke, Rohrer and Yang \cite{hep-ph/0512258} have examined $B \to VV$ decays in the context of QCD factorization.

According to naive factorization, the ratio of longitudinal, negative and positive helicity $B \to VV$ decay amplitudes scale as \cite{hep-ph/0405134}
\begin{equation}
A_0:A_-:A_+ = 1:\frac{\Lambda}{m_b}:\left(\frac{\Lambda}{m_b}\right)^2.
\end{equation}
Although this scaling is well obeyed by $B\to\rho\rho$ decays, it is badly broken by penguin dominated decays such as $B \to K^* \rho$ and $B \to K^* \phi$, which have anomalously large transverse polarizations \cite{hep-ex/0603003}

\begin{figure}
\caption{\label{Q7diags}New contributions to $B \to VV$ decay}
\scalebox{1}[1]{\includegraphics{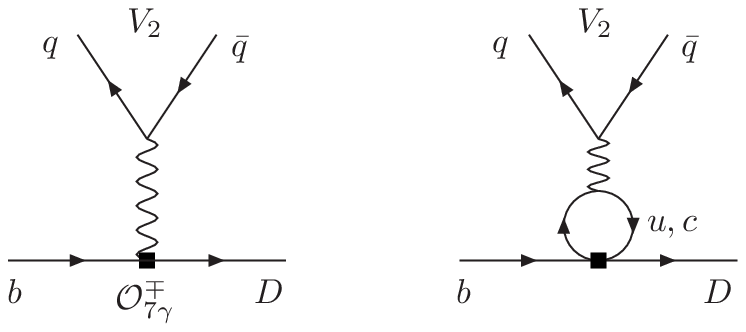}}
\end{figure}

Beneke, Rohrer and Yang have identified a novel contribution to $B \to VV$ decays that had not been previously taken into account.  This contribution, a contribution to the transverse electromagnetic penguin amplitudes, is given by the diagrams in Fig. \ref{Q7diags}. 

Although this contribution is in fact $\mathcal{O}(m_b)$ enhanced over the leading terms, it is also $\alpha_\mathrm{em}$ suppressed.  As a result, it is the dominant contribution to the transverse helicity modes, but is not large enough to
explain the puzzle of the large transverse polarizations of $K^*\rho$ and $K^*\phi$.

Although insufficient to resolve the discrepancy between theory and experiment, the new contribution has pronounced effects, as demonstrated in the examples below.  These examples make use of the current experimental data to estimate the QCD penguin contributions to the decay amplitudes.  QCD factorization is used to estimate all other contributions, including the electroweak penguin effects.  (Several other approximations are also necessary as explained in \cite{hep-ph/0512258}.)

As a first example, consider the ratio of $CP$ averaged helicity decay rates
\begin{equation}
S_h = \frac{2\bar \Gamma_h(\rho^0 \bar K^{*0})}{\bar \Gamma_h(\rho^- \bar K^{*0})}=\left|1-\frac{P_h^{EW}}{P_h}\right|+\Delta_h,
\end{equation}
where $h$ denotes the polarization ($h = 0, -, +$), $P_h^\mathrm{EW}$ is the electroweak penguin contribution calculated in QCDF and $P_h$ is the QCD penguin contribution whose magnitude is determined from the Br and $f_L$ of $\bar B \to \rho^- \bar K^{*0}$ (note that $A_h(\rho^- \bar K^{*0})=P_h$).  $\Delta_h$ is a small contribution from the 
color-suppressed tree amplitude (with some dependence on $P_h^\mathrm{EW}$).  Without the effects of Fig. \ref{Q7diags}, $S_-$ takes on the value $0.7 \pm 0.1$, but is increased to $S_- = 1.5 \pm 0.2$ when the new effects are introduced.

Also consider the ratio
\begin{equation}
S'_h = \frac{2\bar \Gamma_h(\rho^0 \bar K^{*-})}{\bar \Gamma_h(\rho^- \bar K^{*0})}=\left|1-\frac{P_h^\mathrm{EW}}{P_h}\right|+\Delta'_h.
\end{equation}
As before, $|P_h|$ is determined from the Br and $f_L$ for $\bar B \to \rho^- \bar K^{*0}$.
This analysis leads to $S_- = 0.5 \pm 0.1$ with the new contribution and $S_-=1.2 \pm 0.1$ without.

Here it is possible to compare this result to experiment.  The above ratio of branching ratios is related to the ratio of polarization fractions $f'_h \equiv f_h(\rho^0 \bar K^{*-})/f_h(\rho^- \bar K^{*0})$ so long as the $CP$ asymmetry is assumed to be zero.  Using theory to estimate this asymmetry leads to the results
\begin{eqnarray}
f'_0 &=& 1.3 \pm 0.1 \;\;\; [1.1 \pm 0.1 \mbox{(without)}] \nonumber \\
f'_- &=& 0.4 \pm 0.1 \;\;\; [0.8 \pm 0.1 \mbox{(without)}].
\end{eqnarray}
The experimental results for these ratios are
\cite{hep-ex/0307026, hep-ex/0408017, hep-ex/0408093, hep-ex/0307014, hep-ex/0503013, hep-ex/0505039}
$f'_0=1.45^{+0.64}_{-0.58}$ and $f'_-=0.12^{+0.44}_{-0.11}$.
As can be seen, the effects of the new operator can be large (as is the case for $f'_-$) and would seem to bring theory further in line with experimental expectations.

\section{\label{NLO}NLO Corrections in $\alpha_s$ and $1/m_b$}

In this section, we examine current status and outlook for the NLO corrections to the factorization formalism (be it SCET or QCD factorization).  The NLO corrections include both perturbative corrections to the calculable kernels ($T^\mathrm{I}$ and $T^\mathrm{II}$), as well as power corrections ($1/m_b$) to the leading order result Eq. (\ref{LOeqn}).

The primary motivation to include the NLO corrections is simple: to reduce overall uncertainties.  Both perturbative and power corrections are expected to be of order $\sim 20\%$.  
Thus, to reduce theory uncertainties below this level and remain competitive with experiment, it becomes necessary to include both these corrections.

There are also observables for which the NLO corrections are indispensable.
$CP$ asymmetries will receive large corrections at NLO, if at LO the ``tree'' amplitudes (for definition see e.g. \cite{hep-ph/0601214}) do not originate from the $Q_{1,2}$ operators of the effective electroweak Hamiltonian.
For $\Delta S=1$ modes such as $B_s^0 \to \pi^0 \eta^{(\prime)}$ or $B^- \to \bar K^0 \pi^-$ we expect large $\mathcal{O}(1)$ corrections to their $CP$ asymmetries.  This situation is even more extreme for $\Delta S=0$ modes such as $B^- \to K^0 K^-$ or $\bar B^0 \to K^0 \bar K^0$, where no CKM hierarchy exists between the ``tree'' and ``penguin'' amplitudes.

\subsection{Status of NLO corrections}

Typical power corrections are of the order $\sim 20\%$, and will likely introduce many new parameters.  SCET is the theoretical framework necessary to describe the $1/m_b$ corrections, but almost no $1/m_b$ corrections have been considered.   Exceptions include the annihilation contributions and chirally enhanced contributions that have been taken into account by BBNS.  Much work remains to be done for these corrections.

Unlike the power corrections above, almost all of the order $\alpha_s$ corrections have been calculated.  To examine these corrections, it is useful to rewrite the basic QCDF formula (\ref{LOeqn}), by writing $T^{II}=H^{II}*J$,
\begin{eqnarray}
\langle M_1 M_2|O|\bar B\rangle &=& (F^{B\to M_1} T^I+\Xi^{B\to M_1} *H^{II}) \nonumber \\
 &\,&\hspace{1 em}*f_{M_2} \Phi_{M_2} + M_1 \leftrightarrow M_2
\end{eqnarray}
where
\begin{equation}
\Xi^{B\to M_1} \equiv J * f_{B} \Phi_{B}*f_{M_1} \Phi_{M_1}
\end{equation}
$T^\mathrm{I}$ and $T^\mathrm{II}$ are hard coefficients, calculable in $\alpha_s(m_b)\sim0.2$ where perturbation theory is expected to be valid.  The LO contribution to the hard coefficients is in general $\mathcal{O}(1)$.

The jet function, $J$, is perturbatively calculable in $\alpha_s(\mu_i)$ and differs from the hard coefficients in two important ways:  the LO contribution to $J$ is $\mathcal{O}(\alpha_s(\mu_i))$ and $J$ depends on only one of the two final state mesons, and so can be related to other processes.  In particular, the same jet function occurs in semi-leptonic decay.
As of the last FPCP conference, FPCP 2004, the hard kernel
$T^\mathrm{I}$ was known to $\mathcal{O}(\alpha^2_s(m_b) \beta_0)$ (NLO+)
\cite{hep-ph/0101190, hep-ph/0102219, hep-ph/0202128, hep-ph/0504024}, 
$H^\mathrm{II}$ was known to $\mathcal{O}(\alpha^0_s(m_b))$ (LO),
and $J$ was known to $\mathcal{O}(\alpha^2_s(\mu_i))$ (NLO).

\subsection{Corrections to $H^\mathrm{II}$}

Recently a subset of the $\mathcal{O}(\alpha_s(m_b))$ (NLO) corrections to $H^\mathrm{II}$ have been calculated \cite{hep-ph/0512351}.  This calculation involved matching the effective electroweak Hamiltonian onto leading order SCET operators at one loop.  In this calculation, only the operators $Q_{1,2}$ of the electroweak Hamiltonian were matched, leaving the remaining operators for a future work.
Taken with the earlier results for the NLO calculation of the jet function $J$, $\Xi^{B\to M_1}$ became known to NLO for a subset of operators.

As a phenomenological application of these new corrections, the authors re-calculated the predicted $\pi \pi $ branching ratios.
The amplitudes for these decays are given by ($C=i \,\frac{G_F}{\sqrt{2}} m_B^2 f_\pi f^{B\pi}_+(0)$)
\begin{eqnarray}
   \sqrt2\,{\cal A}_{B^-\to\pi^-\pi^0}
   &=&  C
       \left\{ V_{ub} V_{ud}^*\Big[\alpha_1+\alpha_2\Big]\right\}\!, 
   \nonumber\\
   {\cal A}_{\bar B^0\to\pi^+\pi^-}
   &=&  C
       \left\{ V_{ub} V_{ud}^* \Big[\alpha_1 + \hat \alpha_4^u \Big] +
         V_{cb} V_{cd}^* \,\hat \alpha_4^c\right\}\!, 
   \nonumber\\
   -\,{\cal A}_{\bar B^0\to\pi^0\pi^0}
   &=& C 
       \left\{ V_{ub} V_{ud}^* \Big[\alpha_2 
    - \hat\alpha_4^u\Big] - 
         V_{cb} V_{cd}^*\,\hat\alpha_4^c \right\}\!,
         \nonumber 
\end{eqnarray}
where the $\alpha_i$ are the amplitude coefficients defined in \cite{hep-ph/0308039}.
Only the tree amplitudes $\alpha_{1,2}$ are affected by the new contributions; they become,
\begin{eqnarray}
\alpha_1(\pi\pi) &=& 1.015 + [0.025+0.012i\,]_{V}- [0.009]_{\rm tw3} 
\nonumber \\ 
&\,&- \,0.014_\mathrm{LO}- \, [0.024+ 0.020 i\,]_{\rm NLO} 
\nonumber\\
&=& 0.992^{+0.029}_{-0.054}+(-0.007^{+0.018}_{-0.035})\,i,
\end{eqnarray}
\begin{eqnarray}
\alpha_2(\pi\pi) &=& 0.184 - [0.152+0.077i\,]_{V} + [0.056]_{\rm tw3}
\nonumber \\ 
&\,&+ \,0.088_\mathrm{LO} + \, [0.029+ 0.034 i\,]_{\rm NLO} 
\nonumber\\
&=& 0.205^{+0.171}_{-0.110}+(-0.043^{+0.083}_{-0.065})\,i.
\end{eqnarray}
Here the leading order (tree) terms are given no subscript, the NLO ($\mathcal{O}(\alpha_s)$) vertex corrections are denoted with a subscript V, the included subset of power corrections are denoted with the subscript tw3, while LO and NLO denote the LO and NLO (new) spectator scattering corrections.  As can be seen, the new corrections contribute important new strong phases to the amplitudes.

The above numbers offer an important insight into the validity of perturbation theory at the intermediate scale $\sqrt{\Lambda_\mathrm{QCD} m_b}$.  In $\alpha_2(\pi\pi)$, the NLO spectator scattering terms are smaller than the LO terms in agreement with naive expectations.  In $\alpha_1(\pi\pi)$, the NLO corrections appear to be enhanced over the LO term, and may lead one to believe that perturbation theory is breaking down.  This is not a valid conclusion however as the LO term is suppressed as it is proportional to a small Wilson coefficient.  These corrections therefore suggest that perturbation theory is valid at the intermediate scale.

In trying to reproduce the $\pi\pi$ branching ratios (see Table \ref{Brs}), the data favours inputs where
\begin{eqnarray}
|V_{ub}|f_+^{B\pi}(0) &=& 8.10 \times 10^{-4} = 0.775 \left[|V_{ub}|f_+^{B\pi}(0) \right]_\mathrm{def},
\nonumber\\
\frac{f_\pi f_B}{m_b f_+^{B\pi}(0) \lambda_B} &=& 1.96 \left[ \frac{f_\pi f_B}{m_b f_+^{B\pi}(0) \lambda_B}\right]_\mathrm{def},
\end{eqnarray}
with the subscript def. refering to the default values used in the determination of $\alpha_{1,2}(\pi\pi)$ above.
These inputs are consistent with values of the form factor and $\lambda_B$ smaller than their default values \cite{hep-ph/0512351},
\begin{eqnarray}
f_+^{B\pi}(0) &=& 0.22 \;\;\;\;[0.28\pm0.05]_\mathrm{def},\nonumber \\
\lambda_B &=& 0.23 \;\;\;\;[0.35\pm0.15]_\mathrm{def},
\end{eqnarray}
and brings the form factor closer to the BPRS value:  $0.176\pm0.036$ \cite{hep-ph/0401188, hep-ph/0601214}. 
As well, these values remain consistent with what is found using lattice and semi-leptonic data \cite{hep-ph/0509090}.

\begin{table}
\caption{\label{Brs}CP-averaged branching fractions ($\times 10^{-6}$)  Val. denotes the central theory prediction, CKM, hadr. and pow. denote the theory uncertainty associated with CKM matrix elements, hadronic parameters and power corrections respectively, and expt. denotes the present experimental value.}
\begin{tabular}{|c|cccc|c|}
\hline
Mode & Val. & CKM & hadr. & pow. & expt. \\
\hline
$B^-\to \pi^-\pi^0$ & 5.5 & $^{+0.3}_{-0.3}$ & $^{+0.5}_{-0.4}$ & $^{+0.9}_{-0.8}$ & $5.5\pm 0.6$ \\
$\bar{B}^0\to \pi^+\pi^-$ & 5.0 & $^{+0.8}_{-0.9}$ & $^{+0.3}_{-0.5}$ & $^{+1.0}_{-0.5}$ & $5.0\pm 0.4$ \\
$\bar{B}^0\to \pi^0\pi^0$ & 0.7 & $^{+0.3}_{-0.2}$ & $^{+0.5}_{-0.2}$ & $^{+0.4}_{-0.3}$ & $1.5\pm 0.3$ \\
\hline
\end{tabular}
\end{table}

\subsection{At NLO, BBNS - BPRS = ?}

It is instructive to ask what differences remain between the BBNS and BPRS approachs once NLO corrections are taken into account?  At least one of the differences is no longer present at NLO.  In the BPRS approach at LO, the SCET form factor $\Xi^{B\to M_1}$ was fit from $B \to M_1 M_2$ data. 
At NLO one can no longer fit for $\Xi^{B\to M_1}$; there is a non-trivial convolution and not simple moments as at LO.  One
is forced to follow the path of BBNS and expand $\Xi^{B\to M_1}$ in terms of the jet function $J$ as well as other long distance factors.

The question as to whether the soft overlap function should be fit from $B\to MM$ data (BPRS) or be determined by use QCD sum rules or lattice QCD (BBNS) remains unresolved.  BBNS may be moving in the direction of BPRS in that the value of $\lambda$ seems to be determined from fit to $B \to \pi\pi$ modes.

One issue that remains unchanged is what to do with the charming penguins.  For the near future, the disagreement as to whether the charming penguins should be considered to have long distance effects and fit (BPRS) or whether they are assumed perturbative and calculated (BBNS) will remain an open question.  Fitting the charming penguins is always a valid procedure, but lacks the predictive power of perturbative calculation.  A comparison of theoretical predictions between the two approaches should one-day resolve the issue.

\section{Conclusions}

QCD factorization is an important tool in the study of flavour physics.
Reduction of the theoretical uncertainties associated with QCD factorization predictions requires NLO corrections to be computed.
Although the complete set of NLO in $\alpha_s$ corrections is almost complete, only a handful of $1/m_b$ corrections have been considered.

In this work we presented a brief summary of the QCD factorization formalism as well as some recent developments and discussed its similarity to SCET.  We examined several of the differences between the approach put forth by Beneke, Buchalla, Neubert and Sachrajda (BBNS) and the approach advocated by Bauer, Pirjol, Rothstein and Stewart (BPRS).  We concluded that many of the characteristics that differentiate between the two approaches are not applicable when NLO corrections are taken into account.

\bigskip 
\begin{acknowledgments}
The author is indebted to the organizers of FPCP06 for a stimulating and enlightening conference, to Craig Burrell for reviewing the manuscript and to Jure Zupan for invaluable discussion and feedback.  This work was supported in part by the United States Department of Energy under Grants No.\ DOE-ER-40682-143 and DEAC02-6CH03000.
\end{acknowledgments}

\bigskip 

\begin{thebibliography}{99} 

\bibitem{hep-ph/9905312}
  M.~Beneke, G.~Buchalla, M.~Neubert and C.~T.~Sachrajda,
  Phys.\ Rev.\ Lett.\  {\bf 83}, 1914 (1999)
  [arXiv:hep-ph/9905312].

\bibitem{hep-ph/0006124}
  M.~Beneke, G.~Buchalla, M.~Neubert and C.~T.~Sachrajda,
  Nucl.\ Phys.\ B {\bf 591}, 313 (2000)
  [arXiv:hep-ph/0006124].

\bibitem{hep-ph/0104110}
  M.~Beneke, G.~Buchalla, M.~Neubert and C.~T.~Sachrajda,
  Nucl.\ Phys.\ B {\bf 606}, 245 (2001)
  [arXiv:hep-ph/0104110].

\bibitem{hep-ph/0401188}
  C.~W.~Bauer, D.~Pirjol, I.~Z.~Rothstein and I.~W.~Stewart,
  Phys.\ Rev.\ D {\bf 70}, 054015 (2004)
  [arXiv:hep-ph/0401188].

\bibitem{hep-ph/0411171}
  M.~Beneke, G.~Buchalla, M.~Neubert and C.~T.~Sachrajda,
  Phys.\ Rev.\ D {\bf 72}, 098501 (2005)
  [arXiv:hep-ph/0411171].

\bibitem{hep-ph/0502094}
  C.~W.~Bauer, D.~Pirjol, I.~Z.~Rothstein and I.~W.~Stewart,
  Phys.\ Rev.\ D {\bf 72}, 098502 (2005)
  [arXiv:hep-ph/0502094].

\bibitem{hep-ph/0210085}
  M.~Beneke and M.~Neubert,
  Nucl.\ Phys.\ B {\bf 651}, 225 (2003)
  [arXiv:hep-ph/0210085].

\bibitem{hep-ph/0308039}
  M.~Beneke and M.~Neubert,
  Nucl.\ Phys.\ B {\bf 675}, 333 (2003)
  [arXiv:hep-ph/0308039].

\bibitem{hep-ph/0512258}
  M.~Beneke, J.~Rohrer and D.~Yang,
  Phys.\ Rev.\ Lett.\  {\bf 96}, 141801 (2006)
  [arXiv:hep-ph/0512258].

\bibitem{hep-ph/0405134}
  A.~L.~Kagan,
  Phys.\ Lett.\ B {\bf 601}, 151 (2004)
  [arXiv:hep-ph/0405134].

\bibitem{hep-ex/0603003}
  E.~Barberio {\it et al.}  [The Heavy Flavor Averaging Group],
  arXiv:hep-ex/0603003.

\bibitem{hep-ex/0307026}
  B.~Aubert {\it et al.}  [BABAR Collaboration],
  Phys.\ Rev.\ Lett.\  {\bf 91}, 171802 (2003)
  [arXiv:hep-ex/0307026].

\bibitem{hep-ex/0408017}
  B.~Aubert {\it et al.}  [BABAR Collaboration],
  Phys.\ Rev.\ Lett.\  {\bf 93}, 231804 (2004)
  [arXiv:hep-ex/0408017].

\bibitem{hep-ex/0408093}
  B.~Aubert  [BABAR Collaboration],
  arXiv:hep-ex/0408093.

\bibitem{hep-ex/0307014}
  K.~F.~Chen {\it et al.}  [Belle Collaboration],
  Phys.\ Rev.\ Lett.\  {\bf 91}, 201801 (2003)
  [arXiv:hep-ex/0307014].

\bibitem{hep-ex/0503013}
  K.~F.~Chen {\it et al.}  [BELLE Collaboration],
  Phys.\ Rev.\ Lett.\  {\bf 94}, 221804 (2005)
  [arXiv:hep-ex/0503013].

\bibitem{hep-ex/0505039}
  J.~Zhang {\it et al.}  [BELLE Collaboration],
  arXiv:hep-ex/0505039.

\bibitem{hep-ph/0601214}
  A.~R.~Williamson and J.~Zupan,
  arXiv:hep-ph/0601214.

\bibitem{hep-ph/0101190}
  C.~N.~Burrell and A.~R.~Williamson,
  Phys.\ Rev.\ D {\bf 64}, 034009 (2001)
  [arXiv:hep-ph/0101190].

\bibitem{hep-ph/0102219}
  T.~Becher, M.~Neubert and B.~D.~Pecjak,
  Nucl.\ Phys.\ B {\bf 619}, 538 (2001)
  [arXiv:hep-ph/0102219].

\bibitem{hep-ph/0202128}
  M.~Neubert and B.~D.~Pecjak,
  JHEP {\bf 0202}, 028 (2002)
  [arXiv:hep-ph/0202128].

\bibitem{hep-ph/0504024}
  C.~N.~Burrell and A.~R.~Williamson,
  arXiv:hep-ph/0504024.

\bibitem{hep-ph/0512351}
  M.~Beneke and S.~Jager,
  arXiv:hep-ph/0512351.

\bibitem{hep-ph/0509090}
  T.~Becher and R.~J.~Hill,
  Phys.\ Lett.\ B {\bf 633}, 61 (2006)
  [arXiv:hep-ph/0509090].

\end{thebibliography}

\end{document}